\documentstyle[epsfig,twocolumn,prl,aps]{revtex}
\begin{document}
\draft
\twocolumn[\hsize\textwidth\columnwidth\hsize\csname @twocolumnfalse\endcsname
\title{Entanglement Teleportation via Werner States} \author{Jinhyoung
  Lee$^1$ and M. S.
  Kim$^{1,2}$} 
\address{$^1$ Department of
  Physics, Sogang University, CPO Box 1142, Seoul 100-611, Korea
\\
  $^2$ Department of Applied Mathematics and Theoretical Physics, 
\\ The Queen's University of Belfast, BT7 1NN, UK }
\date{\today} \maketitle

\begin{abstract}  
  Transfer of entanglement and information is studied for quantum
  teleportation of an unknown entangled state through noisy quantum
  channels. We find that the quantum entanglement of the unknown state
  can be lost during the teleportation even when the channel is
  quantum correlated.  We introduce a fundamental parameter of {\em
    correlation information} which dissipates {\em linearly} during
  the teleportation through the noisy channel.  Analyzing the transfer
  of correlation information, we show that the purity of the initial
  state is important in determining the entanglement of the replica
  state.
\end{abstract}

\vskip1pc] 

\newpage

The nonlocal property of quantum mechanics enables a striking
phenomenon called quantum teleportation.  By quantum teleportation
an unknown quantum state is destroyed at a sending place while its perfect
replica state appears at a remote place via dual quantum and classical
channels \cite{bennett93,Zeilinger98}.  For the perfect quantum
teleportation, a maximally entangled state, {\it e.g.} a singlet state,
is required for the quantum channel.  However, the decoherence effects
due to the environment make the pure entangled state into a
statistical mixture and degrade quantum entanglement in the real
world.  Popescu \cite{popescu94} studied the quantum teleportation
with the mixed quantum channel and found that even when the channel is
not maximally entangled, it has the fidelity better than any classical
communication procedure.  For a practical purpose, a purification
scheme may be applied to the noisy channel state before teleportation.
\cite{bennett96,linden98,horodecki97}.

Earlier studies have been confined to the teleportation of single-body
quantum states: Quantum teleportation of two-level states
\cite{bennett93}, $N$-dimensional states \cite{Stenholm98}, and
continuous variables \cite{Vaidman94,Ralph98}.  In this Letter, we are
interested in teleportation of two-body entangled quantum states,
especially regarding the effects of the noisy environment. Direct
transmission of an entangled state was considered in a noisy
environment \cite{schumacher96}.  A possibility to copy pure entangled
states was studied \cite{Koashi98}.  Extending the argument of the
single-body teleportation we can easily show that an entangled
$N$-body state can be perfectly teleported using the $N$
maximally-entangled pairs for the quantum channel.  However, for the
noisy channel, it is important and nontrivial to know how much the
entanglement is transferred to the replica state and how close the
replica state is to the original unknown state, depending on the
entanglement of the unknown state and channel state.  

Bennett {\it et
  al.} \cite{bennett93} argued that teleportation is a linear
operation for the perfect quantum channel so that it would also work
with mixed states and could be extended to what is now called
entanglement swapping \cite{Zukowski93}.  We rigorously found that
teleportation is linear even for the mixed channel, considering the
maximization of the average fidelity \cite{Lee-Sogang99}.  With the
property of the linearity, one may {\em conjecture} that quantum
teleportation preserves the nature of quantum correlation in the
unknown entangled state if the channel is quantum-mechanically
correlated. We investigate this conjecture.

In this Letter, the original unknown state is assumed to be in an
entangled two-body pure spin-1/2 state and the noisy quantum channel
to be represented by a Werner state \cite{werner89}.
We define the measure of entanglement for the two spin-1/2 system and
study the transfer of entanglement in the teleportation.  We find that
for the quantum channel there is a {\it critical value of minimum
  entanglement} required to teleport quantum entanglement. This
minimum entanglement is understood by considering the transfer of
entanglement and correlation information. 
The newly-defined correlation
information, which dissipates
{\em linearly} during the teleportation through the noisy channel, is
related to quantum entanglement for a pure state, and may also be to
classical correlation for a mixed state.  Analyzing the
transfer of correlation information, it is shown that the {\it purity of
  the initial state} is important in determining the entanglement of
the replica state.

Before considering the entanglement teleportation procedure, we define
a measure of entanglement. Consider a density matrix
$\hat{\rho}$ and its partial transposition
$\hat{\sigma}= \hat{\rho}^{T_2}$ for a two spin-1/2 system.  The density
matrix $\hat{\rho}$ is inseparable if and only if $\hat{\sigma}$ has
any negative eigenvalues \cite{peres96_1,horodecki96_2}. The measure
of entanglement ${\cal E}(\hat{\rho})$ is then defined by
\begin{equation}
  \label{eq:moe}
  {\cal E}(\hat{\rho}) = - 2\sum_i \lambda^-_i
\end{equation}
where $\lambda^-_i$ is a negative eigenvalue of $\hat{\sigma}$.
It is straightforward to prove that ${\cal E}(\hat{\rho})$ satisfies
the necessary conditions required for every measure of entanglement
\cite{Lee-Sogang99,vedral9798}.

The entanglement teleportation is schematically plotted in Fig.~1.
Sender's unknown state $\hat{\rho}_{12}$ is prepared by the source
$S$.  Two {\it independent} EPR pairs are generated from $E$ (one pair
numbered 3 and 5, the other pair 4 and 6 in Fig.~1).  When a noisy
environment is considered, its effects are attributed to the quantum
channels and the perfect EPR pair becomes mixed.  By applying
random $SU(2)$ operations locally to both members of a pair a general
mixed two-body state becomes a highly symmetric Werner state which is
$SU(2)\otimes SU(2)$ invariant \cite{bennett96,werner89}. For example,
the quantum channel $Q_1$ is represented by the density matrix
$\hat{w}_{35}$ of purity $(\Phi_{35}+1)/2$ \cite{werner89}:
\begin{equation}
  \label{eq:qc}
  \hat{w}_{35} = \frac{1}{4}\left(I\otimes I - \frac{2\Phi_{35}+1}{3}
  \sum_{n} \sigma_n\otimes \sigma_n\right)
\end{equation}
where $\sigma_n$ is a Pauli matrix.  The parameter $\Phi_{35}$ is
related to the measure of entanglement ${\cal E}_{35}$, i.e., ${\cal
  E}_{35} \equiv {\cal E}(\hat{w}_{35}) = \mbox{max}(0, \Phi_{35})$.
To make our discussion simpler, we assume that the two independent
quantum channels are equally entangled, i.e., ${\cal E}_{35} = {\cal
  E}_{46} \equiv {\cal E}_w$.  This assumption can be justified as the
two quantum channels are influenced by the same environment.

At $A_i$, a Bell-state measurement is performed on the particle $i$
from $S$ and one of the pair, $i+2$, in the quantum channel $Q_i$.
The Bell-state measurement at $A_i$ is then represented by a family of
projectors $\hat{P}_i^\alpha = |\Psi_i^\alpha\rangle
\langle\Psi_i^\alpha|$ with $\alpha=1,2,3,4$, where
$|\Psi_i^\alpha\rangle$ are the four possible Bell states.  The joint
measurements at $A_1$ and $A_2$ project the total density matrix
$\hat{\rho}$ on to the Bell states $|\Psi_1^\alpha\rangle$ and
$|\Psi_2^\beta\rangle$, respectively, with the probability
$P_{\alpha\beta} ={\rm Tr}\hat{P}_1^\alpha\hat{P}_2^\beta \hat{\rho}$.
The probability $P_{\alpha\beta}$ is 1/16 which is a characteristic of
the Werner state.  After receiving the two-bit information on the
measurements through the classical channels $C_1$ and $C_2$, the
unitary transformations $\hat{U}^\alpha_1$ and $\hat{U}^\beta_2$ are
performed on the particles 5 and 6 accordingly.
 
By the unitary transformations, we reproduce the unknown state at
$B_1$ and $B_2$ if the channel is maximally entangled.  In choosing
$\hat{U}_i^\alpha$, an important parameter to consider is the
fidelity ${\cal F}$ defined as the distance between the unknown pure
state $\hat{\rho}_{12}$ and the replica state $\hat{\rho}_{78}$:
${\cal F} = {\rm Tr}\hat{\rho}_{12}\hat{\rho}_{78}$.
If $\hat{\rho}_{78}=\hat{\rho}_{12}$ then ${\cal F}=1$. It shows that
the replica is exactly the same as the unknown state and the
teleportation has been perfect.  The four unitary operations are
given by the Pauli spin operators for the singlet-state channel:
$\hat{U}^1_i = \hat{1}, ~\hat{U}^2_i = \hat{\sigma}_x,
  ~\hat{U}^3_i = \hat{\sigma}_y, ~\hat{U}^4_i = \hat{\sigma}_z$.
For the Werner-state channel, we found that the same set of unitary
operations $\hat{U}^\alpha_i$ are applied to maximize the fidelity
\cite{Lee-Sogang99}.  The density matrices of 
both the original unknown state and the replica
state can be written in the same form:  
\begin{equation}
  \label{eq:rhop}
  \hat{\rho} = \frac{1}{4}\left(I\otimes I +
  \vec{a}\cdot\vec{\sigma}\otimes I + 
  I\otimes \vec{b}\cdot\vec{\sigma} +
  \sum_{nm}c_{nm}\sigma_n\otimes\sigma_m\right).
\end{equation}
The real vectors $\vec{a}$, $\vec{b}$, and real matrix $c_{nm}$
of the replica state $\hat{\rho}_{78}$ is related with 
$\vec{a}_0$, $\vec{b}_0$, and $c_{nm}^0$ of the original state:
$\vec{a}=(2\Phi_w+1)\vec{a}_0/3$,
$\vec{b}=(2\Phi_w+1)\vec{b}_0/3$, and
$c_{nm}=(2\Phi_w+1)(2\Phi_w+1)c^0_{nm}/9$.

The maximum fidelity ${\cal F}$ depends on the initial entanglement
${\cal E}_{12}={\cal E}(\hat{\rho}_{12})$:
\begin{equation}
{\cal F}={\cal F}^c + {\cal F}^q {\cal E}_{12}^2
\label{eq:fidelity-ent}
\end{equation}
where ${\cal F}^c= (E_w+2)^2/9$, ${\cal F}^q=(2{\cal E}_w+1)({\cal
  E}_w-1)/9$. When the unknown pure
state is not entangled, i.e. ${\cal E}_{12}=0$, the fidelity is just
${\cal F}^c$ which is the maximum fidelity for double teleportation
of independent two particles
\cite{popescu94,gisin96}. For a given channel entanglement, 
the fidelity ${\cal F}$ decreases
monotonously as the initial entanglement ${\cal E}_{12}$ increases
because ${\cal F}^q \le 0$. To obtain the same fidelity, the
larger entangled channels are required for the larger entangled
initial state.  It implies that the entanglement is so
fragile to teleport.

The measure of entanglement ${\cal  E}_{78}$ for the 
replica state $\hat{\rho}_{78}$ is found using its definition 
(\ref{eq:moe}) as
\begin{equation}
  \label{eq:entrp}
  {\cal E}_{78} = \mbox{max}\left\{0, \frac{1}{9} \left[(2{\cal
  E}_w^2+2{\cal E}_w-4) + 
  (1+2{\cal E}_w)^2 {\cal E}_{12}\right]\right\}.
\end{equation}
In Fig.~\ref{fig:entangle}, the entanglement ${\cal
  E}_{78}$  is plotted with respect to
the entanglement ${\cal E}_{12}$ for the unknown state and ${\cal
  E}_w$ for the quantum channel. We find that 
${\cal E}_{78}$ is nonzero showing entanglement in the replica
state only when ${\cal E}_w$ is larger than a critical value 
${\cal E}^c_w
\equiv(3-\sqrt{2{\cal E}_{12}+1})/(2\sqrt{2{\cal E}_{12}+1})$. 
If the unknown state is maximally
entangled with ${\cal E}_{12}=1$, the quantum channel is required to have
the entanglement larger than ${\cal E}_w^c \sim 0.3660$. It is
remarkable that the entanglement teleportation has the critical value
of minimum entanglement ${\cal E}^c_w \ne 0$ for the quantum channel
to transfer any entanglement.

Brukner and Zeilinger \cite{Brukner99} recently introduced a new measure
of quantum information which is normalized to have $n$ bits of information
for $n$ qubits. Based on their derivation, we define a measure of
correlation information. 
The measure of total information for
the density matrix $\hat{\rho}$ of the two spin-1/2 particles is
${\cal I}(\hat{\rho}) = \frac{2}{3} \left(4{\rm Tr}\hat{\rho}^2 - 
1\right)$,
which may be decomposed into three parts.  Each particle has
its own information  corresponding to its marginal density matrix,
which we call the {\it individual information}. The two particles can
also share the {\it correlation information} which depends
on how much they are correlated.  
The measure of individual information ${\cal I}^a(\hat{\rho})$ 
for the particle $a$ is
\begin{equation}
 \label{eq:oqia}
{\cal I}^a(\hat{\rho}) = 2 {\rm Tr}_a 
\left(\hat{\rho}_a\right)^2 - 1
\end{equation}
where $\hat{\rho}_a = {\rm Tr}_b \hat{\rho}$ is the marginal
density matrix for particle $a$.  The measure of individual information 
${\cal I}^b(\hat{\rho})$ for
particle $b$ can be obtained analogously.  If the total density
matrix $\hat{\rho}$ is represented by $\hat{\rho}= \hat{\rho}_a
\otimes \hat{\rho}_b$, the total system has no correlation.  We define the
measure of correlation information as \cite{corrl}
\begin{equation}
 \label{eq:jqim}
{\cal I}^c(\hat{\rho}) = {\cal I}(\hat{\rho}) - {\cal
 I}(\hat{\rho}_a\otimes\hat{\rho}_b) 
\end{equation}
If there is no correlation between the two particles, the measure of
total information is a mere sum of individual
information.  On the other hand, the total information is imposed only
on the correlation information, ${\cal I}={\cal I}^c$, if there is no
individual information as for the singlet state.  For a two-body
spin-1/2 system, 1 bit is the maximum degree of each individual
information while the correlation information can have maximum 2 bits.

The correlation information is in general contributed from 
quantum entanglement and classical correlation. 
When a pure entangled
state is considered, its entanglement contributes to the whole of
correlation information. For a mixed state, on the other hand, the correlation
information may also be due to classical correlation.  For example, the
Werner state with the entanglement ${\cal E}$ has the correlation
information ${\cal I}^c =\alpha + \beta {\cal E} + \gamma {\cal E}^2$
with constants $\alpha$, $\beta$, and $\gamma$.

The entanglement teleportation transfers the correlation information
${\cal I}^c_{12}\equiv {\cal I}^c(\hat{\rho}_{12}) $ of the unknown state
$\hat{\rho}_{12}$ to the replica state $\hat{\rho}_{78}$.  After a
straightforward algebra, we find that the
transferred correlation information ${\cal I}^c_{78}$ is given by
\begin{equation}
\label{eq:cilt}
{\cal I}^c_{78} = \kappa^4 {\cal I}^c_{12}~~,~~
\kappa={2{\cal E}_w+1 \over 3}
\end{equation}
which  shows that
the correlation information dissipates {\it linearly}  
during the teleportation via the noisy quantum
channel. As far as the channel is entangled
(${1 \over 3} < \kappa \le 1$), some
correlation information remains in the replica state.
Substituting Eq.~(\ref{eq:rhop}) into  
Eq.~(\ref{eq:jqim}), 
we find that the replica
state can have both classical and quantum correlation.  Further, if the
channel is entangled less than ${\cal E}^c_w$, ${\cal I}^c_{78}$
is totally determined by classical correlation. 
The reason why the teleportation does not 
necessarily transfer the entanglement to the replica state 
is that the correlation 
information for the replica state can be determined not only by 
quantum entanglement but also by classical correlation.
We analyze it further as we separate the full teleportation into
two partial teleportations of entanglement. 

Consider a series of two partial
teleportations of entanglement \cite{partial}. After the 
teleportation of particle 1 of the state $\hat{\rho}_{12}$, particle 2
of $\hat{\rho}_{72}$ is teleported and the final replica state is
$\hat{\rho}_{78}$ in Fig.~\ref{fig:configuration}. 
We calculate the transfer of correlation information for the
two teleportations
\begin{equation}
\label{inf-two}
{\cal I}^c_{72} = \kappa^2 {\cal I}^c_{12}~~,~~
{\cal I}^c_{78} = \kappa^2 {\cal I}^c_{72}.
\end{equation}
From these linear equations, we can easily recover 
Eq.~(\ref{eq:cilt}).  
Now we investigate the dependence of correlation
information on entanglement and classical correlation.
For the entangled channel, ${\cal E}_w \ne 0$, the
correlation information ${\cal I}^c_{72}$ can be written in terms of the
entanglement ${\cal E}_{72}$ for $\hat{\rho}_{72}$:
\begin{equation}
  {\cal I}^c_{72} = 2 \kappa^2 \left( 4- 3 
  \frac{{\cal E}_{72}+(1-{\cal E}_w)}{{\cal E}_w(2+{\cal E}_w)} {\cal
  E}_{72}\right)  \frac{{\cal E}_{72}+(1-{\cal E}_w)}{{\cal E}_w(2+{\cal
  E}_w)} {\cal E}_{72}  
\end{equation}
which shows that for ${\cal E}_w
\ne 0$ the correlation information of the state $\hat{\rho}_{72}$ 
is due only to entanglement.  The partial  teleportation $\hat{\rho}_{12}
\rightarrow \hat{\rho}_{72}$ transfers at least some of 
the initial entanglement as far as the
channel is entangled.  However, we have already seen that
the final replica state $\hat{\rho}_{78}$ may include some classical
correlation.  The partial teleportation $\hat{\rho}_{72}
\rightarrow \hat{\rho}_{78}$ may bring about no entanglement transfer.
Why?  The only difference of the two procedures is the purity of
their initial states as $\hat{\rho}_{12}$ is pure while
$\hat{\rho}_{72}$ may be mixed.  
The purity of $\hat{\rho}_{72}$  is
determined by the entanglement of the channel $Q_1$.

To analyze the importance of the initial purity for the entanglement
transfer in partial teleportation, we release, for a while, the hereto
assumption that both the quantum channels have the same measure of
entanglement. The entanglement ${\cal E}_{78}$ for the replica state
then depends on the entanglement ${\cal E}_{46}$ of the quantum
channel $Q_2$, and entanglement ${\cal E}_{72}$ and purity ${\cal
  P}_{72}$ of the state $\hat{\rho}_{72}$.  The more $Q_1$ is
entangled, the purer $\hat{\rho}_{72}$ is.  We numerically calculate
the dependence of entanglement ${\cal E}_{78}$ on the purity ${\cal
  P}_{72}$ of the intermediate state $\hat{\rho}_{72}$ as shown in
Fig.~3.  It clearly shows that the purity of the initial state
determines the possibility of the entanglement transfer.  This
analysis can be analogously applied to the other sequence of partial
teleportations $\hat{\rho}_{12}\rightarrow\hat{\rho}_{18}$ and
$\hat{\rho}_{18}\rightarrow\hat{\rho}_{78}$.

In conclusion, we investigated the effects of the noisy environment on
the entanglement and information transfer in the entanglement teleportation.
The introduction of the measures of entanglement and correlation
information enables us to analyze intrinsic properties of the
entanglement teleportation. We found that the teleportation always
transfers the correlation information which dissipates linearly through
the impure quantum channel. On the other hands, the entanglement
transfer is not always possible. The analysis of partial teleportation 
shows that the purity of an initial state determines the
possibility of the entanglement transfer. We explained this nontrivial
feature by showing that a mixed state can have simultaneously quantum
and classical correlations. Our studies on the entanglement transfer
in the noisy environment will contribute to the entanglement
manipulation, one of basic schemes in quantum information theory.

JL thanks Inbo Kim and Dong-Uck Hwang for 
discussions. This work is supported by the Brain Korea 21 project
of the Korean Ministry of Education.

\begin{figure}
  \begin{center}
    \caption{Schematic drawing of entanglement teleportation. An
      unknown quantum entangled state is generated
      by the source $S$ and its particles are distributed separately
      into $A_1$ and $A_2$. The
      quantum channels $Q_1$ and $Q_2$ are represented by Werner
      states.  The result of the Bell-state
      measurement at $A_i~(i=1,2)$ is transmitted
      through the classical channels $C_i$.  The
      teleportation is completed by unitarily transforming at $B_i$
      according to the classical information.  }
    \label{fig:configuration}
  \end{center}
\end{figure}

\begin{figure}
  \begin{center}
    \caption{Measure of entanglement ${\cal E}_{78}$ for the replica state
      $\hat{\rho}_{78}$ with respect to the entanglement ${\cal
        E}_{12}$ for the unknown pure state and ${\cal E}_w$ for the
      quantum channel.  }
    \label{fig:entangle}
  \end{center}
\end{figure}

\begin{figure}
  \begin{center}
    \caption{For the partial teleportation
      $\hat{\rho}_{72} \rightarrow \hat{\rho}_{78}$ with the channel
      entanglement ${\cal E}_{46}=0.6$, the measure of
      entanglement ${\cal E}_{78}$ for the replica state is plotted
      against the purity ${\cal P}_{72}$. The entanglement ${\cal
      E}_{72}$=0.16 (solid), 0.18 (dotted), 0.20 (dashed), and 0.21
      (long-dashed).}
  \end{center}
\end{figure}


\begin{references} 
  
\bibitem{bennett93} C. H.  Bennett, G. Brassard, C. Cr\'{e}peau, R.
  Jozsa, A. Peres, and W. K. Wootters, Phys. Rev. Lett. {\bf 70}, 1895
  (1993).
  
\bibitem{Zeilinger98} D.  Bouwmester, J-W. Pan, K. Mattle, M. Eibl,
  H.  Weinfurter, and A. Zeilinger, Nature (London) {\bf 390}, 575
  (1997); D. Boschi, S. Branca, F. De Martini, L. Hardy, and S.
  Popescu, \prl {\bf 80}, 1121 (1998).
  
\bibitem{popescu94} S. Popescu, Phys. Rev. Lett. {\bf 72}, 797(1994).
  
\bibitem{bennett96} C. H. Bennett, G. Brassard, S. Popescu, B.
  Schumacher, J. A. Smolin, and W. Wooters, Phys. Rev. Lett. {\bf 76},
  722(1996)
  
\bibitem{linden98}N. Linden, S. Massar, and S. Popescu, \prl {\bf 81},
  3279 (1998).
  
\bibitem{horodecki97} M. Horodecki, P. Horodecki, and R. Horodecki,
  Phys. Rev. Lett. {\bf 78}, 574(1997).
  
\bibitem{Stenholm98} S. Stenholm and P. J. Bardroff, \pra {\bf 58},
  4373 (1998).
  
\bibitem{Vaidman94} L. Vaidman, \pra {\bf 49}, 1473 (1994); S. L.
  Braunstein and H. J. Kimble, \prl {\bf 80}, 869 (1998).
  
\bibitem{Ralph98} T. C. Ralph and P. K. Lam, \prl {\bf 81}, 5668
  (1998); A. Furusawa, J. L. Sorensen, S. L. Braunstein, C. A. Fuchs,
  H. J. Kimble, and E. S. Polzik, Science {\bf 282}, 706 (1998).
 
\bibitem{schumacher96} B. Schumacher, \pra {\bf 54}, 2614 (1996).
  
\bibitem{Koashi98} M. Koashi and N. Imoto, \prl {\bf 81}, 4264 (1998).
  
\bibitem{Zukowski93} M. Zukowski, A. Zeilinger, M. A. Horne, and A.
  Ekert, \prl {\bf 71}, 4287 (1993); J.-W. Pan, D. Bouwmeester, H.
  Weinfurter, and A. Zeilinger, \prl {\bf 80}, 3891 (1998).
  
\bibitem{Lee-Sogang99} J. Lee, Ph.D. thesis, Sogang Univesity (1999).
  
\bibitem{werner89} R. F. Werner, Phys. Rev. A {\bf 40}, 4277(1989).
  
\bibitem{peres96_1} A. Peres, Phys. Rev.  Lett. {\bf 77}, 1413(1996).
  
\bibitem{horodecki96_2} M. Horodecki, P. Horodecki, and R. Horodecki,
  Phys. Lett. A{\bf 223}, 1(1996).
  
\bibitem{vedral9798} V. Vedral, M. B. Plenio, M. A. Rippin, and P. L.
  Knight, Phys. Rev. Lett. {\bf 78}, 2275 (1997); V. Vedral and M. B.
  Plenio, Phys. Rev. A{\bf 57}, 1619 (1998) and references therein.

\bibitem{gisin96} N. Gisin, Phys. Lett.  A{\bf 210}, 157(1996).
  
\bibitem{Brukner99} \v{C}. Brukner and A. Zeilinger, Phys. Rev. Lett.
  {\bf 83}, 3354 (1999).
  
\bibitem{corrl} The measures of individual and correlation information 
  are invariant for any choice of the complete set of complementary
  observables.

\bibitem{partial} Partial teleportation of entanglement is the same
  as entanglement swapping when the quantum channel is maximally entangled.
  
\end{references}
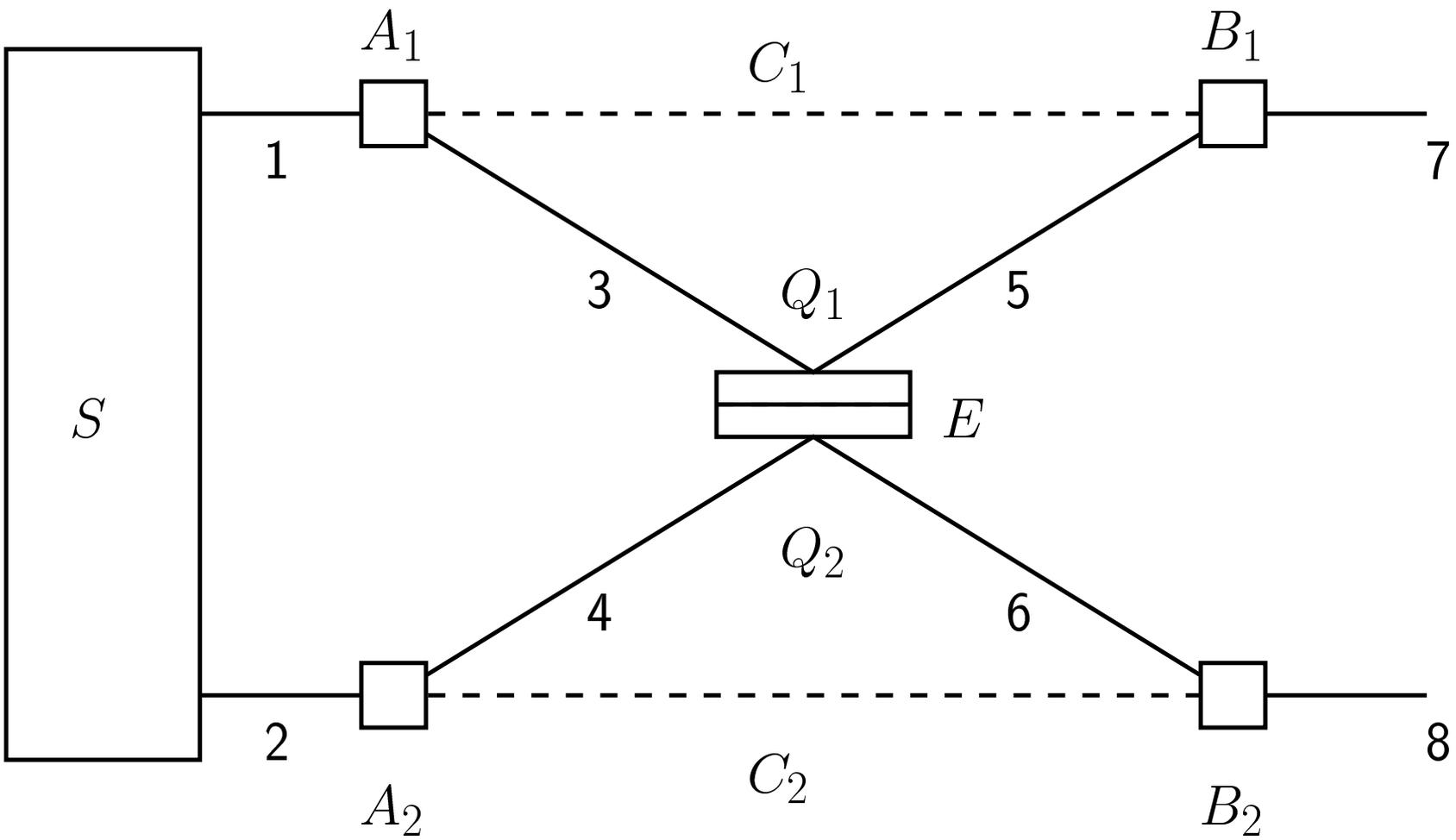
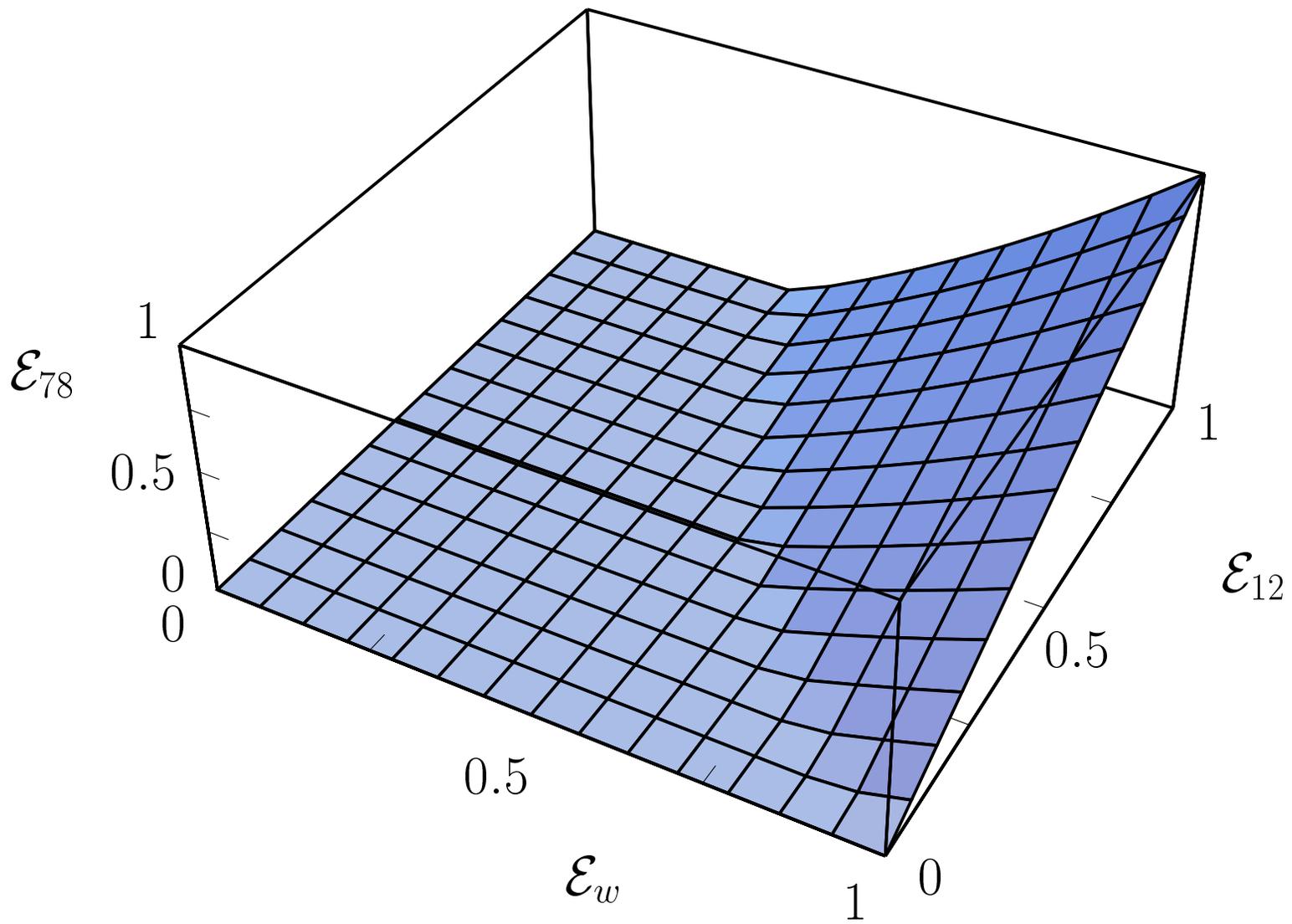
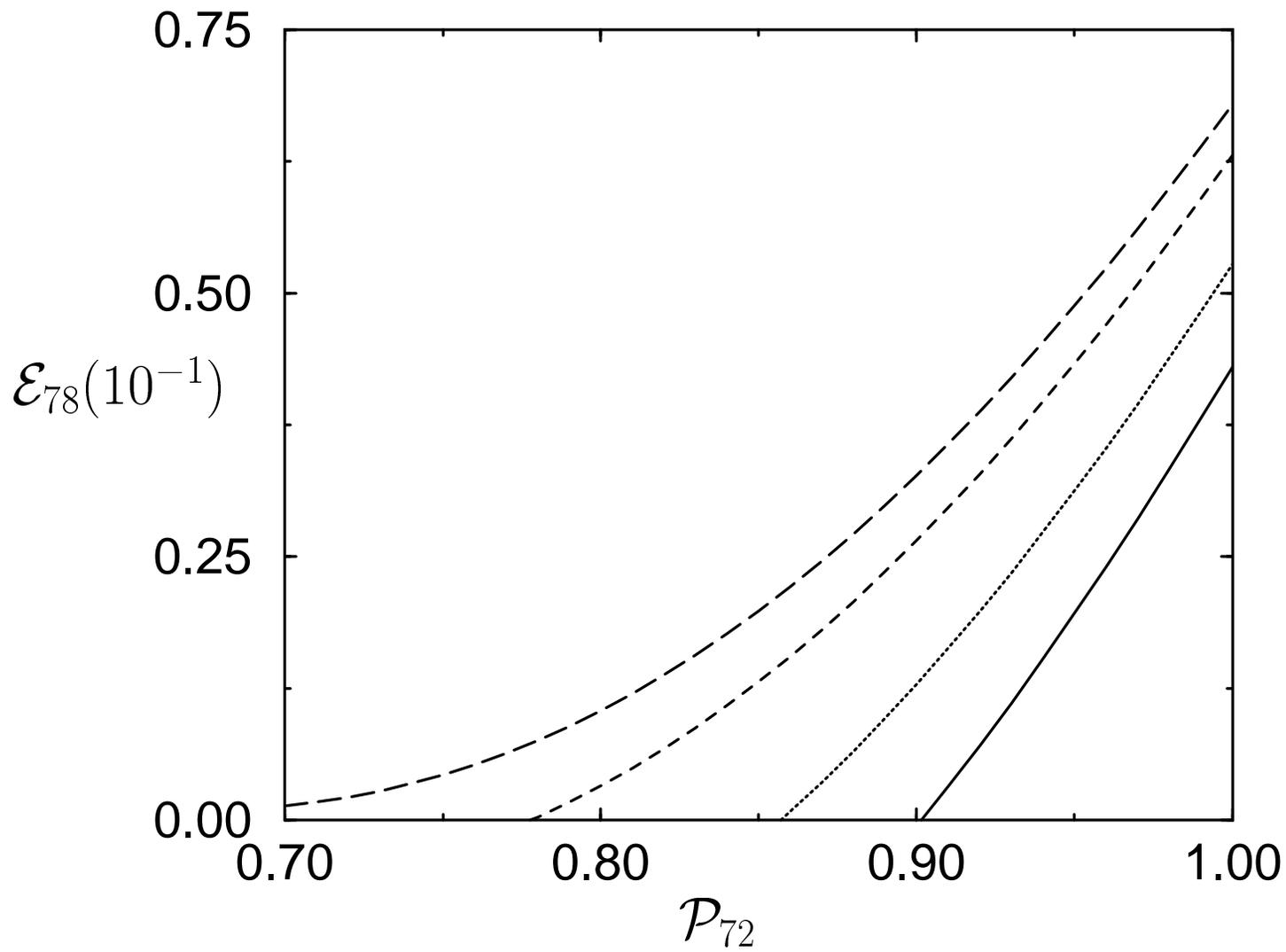
\end{document}